\documentclass[fleqn,usenatbib]{mnras}
\usepackage{newtxtext,newtxmath}
\usepackage[T1]{fontenc}
\usepackage{ae,aecompl}
\usepackage{graphicx}	        
\usepackage{amsmath}	       
\usepackage{url}

\usepackage{caption}
\usepackage{float}
\usepackage{verbatim}
\usepackage[dvipsnames]{xcolor}

\newcommand{\msun}{\ensuremath{\mathit{M}_{\odot}}}   % solar mass
\newcommand{\rsun}{\ensuremath{\mathit{R}_{\odot}}}   % solar radius
\newcommand{\kms}{\ensuremath{{\rm km\,s^{-1}}}}      % kilometers per second
\newcommand{\lsun}{\ensuremath{\mathit{L}_{\odot}}}   % solar luminosity
\newcommand{\msunyr}{\ensuremath{\mathit{M}_{\odot}{\rm yr}^{-1}}} % solar mass per year
\newcommand{\lsn}{\ensuremath{\mathit{L}_{\rm SN}}}   % supernova luminosity
\newcommand{\rin}{\ensuremath{\mathit{R}_{\rm in}}}   % inner radius
\newcommand{\vinf}{\ensuremath{\upsilon_{\infty}}}    % terminal velocity
\newcommand{\mdot}{\ensuremath{\dot{M}}}              % mass-loss rate
\newcommand{\tstar}{\ensuremath{\mathit{T}_{\tau=10}}} % effective temperature

\newcommand{\object}{\textrm{SN~2016bkv}} %object in this paper
\title[The progenitor of SN~2016bkv]{The origins of low-luminosity supernovae: the case of SN~2016bkv}

\author[M. Deckers et al.]{
Maxime Deckers$^{1}$, 
Jose H. Groh$^{1}$\thanks{Contact e-mail: \href{mailto:jose.groh@tcd.ie}{groh.astro@mailbox.org}},
 Ioana Boian$^{1}$,
 Eoin J. Farrell$^{1}$
\\
$^{1}$ Trinity College Dublin, the University of Dublin, Dublin, Republic of Ireland.
}

\date{Accepted 2021 August 19. Received 2021 August 19; in original form 2020 August 1}

\pubyear{2020}

\begin{document}
\label{firstpage}
\pagerange{\pageref{firstpage}--\pageref{lastpage}}
\maketitle
 
\begin{abstract}
We investigate the low-luminosity supernova SN~2016bkv and its peculiar early-time interaction. For that, we compute radiative transfer models using the CMFGEN code. Because SN~2016bkv shows signs of interaction with material expelled by its progenitor, it offers a great opportunity to constrain the uncertain evolutionary channels leading to low-luminosity supernovae. Our models indicate that the progenitor had a mass-loss rate of $(6.0 \pm 2.0)\times10^{-4}$ \msunyr (assuming a velocity of 150~\kms). The surface abundances of the progenitor are consistent with solar contents of He and CNO. If SN~2016bkv's progenitor evolved as a single star, it was an odd red supergiant that did not undergo the expected dredge up for some reason. We propose that the progenitor more likely evolved through binary interaction. One possibility is that the primary star accreted unprocessed material from a companion and avoided further rotational and convective mixing until the SN explosion.  Another possibility is a merger with a lower mass star, with the primary remaining with low N abundance until core collapse. Given the available merger models, we can only put a loose constraint on the pre-explosion mass around 10--20~\msun, with lower values being favored based on previous observational constraints from the nebular phase.
\end{abstract}
%________________________________________________________________

\section{Introduction}
This paper deals with the overarching topic of connecting supernovae (SN) and other transients to their progenitors, with a particular focus on SN 2016bkv. This peculiar event had a significantly lower SN luminosity than average. Linking SN to their progenitors has relevance to many fields of Astrophysics, and yet no comprehensive view exists. This is particularly relevant because the majority of massive stars in the local universe finish their evolution with a SN explosion \citep[e.g.,][]{maeder2000,woosley2002,heger03,langer12}. The vast majority of SN explosions from massive-star progenitors show hydrogen lines in their spectrum, being classified as SN II \citep{Filippenko1997}. Among them, the most common subclass is labeled as SN II-P (plateau), owing to the H recombination plateau observed in their optical lightcurves \citep[e.g.][]{Filippenko1997,arcavi12,anderson14, valenti16}. Stellar evolution models generally predict that the SN II-P progenitors are red supergiants (RSG; e.g. \citealt{heger03,ekstrom2012,groh2013,Choi2016,eldridge17,zapartas19}), and pre-explosion imaging of SN II-P have confirmed RSGs as SN II-P progenitors \citep[e.g.][]{smartt09,smartt15,vandyk12,vandyk12_12aw}. The only blue supergiant (BSG) with pre-explosion observations is SN 1987A's progenitor \citep{walborn89}, which gave rise to a peculiar SN II lightcurve \citep{arnett89}.

Massive stars generally suffer substantial mass loss during their evolution, and sometimes also near core collapse \citep[e.g.][]{smith14araa}. This nicely reflects in the SN spectral morphology, with some SNe exhibiting narrow emission lines in their spectra due to interaction between the SN ejecta and material expelled by the progenitor before explosion. These events are broadly referred to as interacting supernovae, with the SN IIn nomenclature often reserved for long-lived version of these events \citep{schlegel90,Filippenko1997}. If the SN progenitor expels enough material through its winds and/or outbursts in the final years of its life, the SN ejecta will be decelerated by this material, and a fraction of its kinetic energy will be converted to radiation, ionising the rest of the circumstellar material (CSM). The CSM becomes opaque and the photosphere is located out in the CSM, and not in the SN ejecta. Therefore a spectrum taken during this stage will reveal the CSM properties, i.e. the low velocities of the progenitors' wind ($10-2000$ \kms) as opposed to the fast SN ejecta ($\approx 10\,000$ \kms), the progenitors' mass-loss rate, and its surface abundances \citep{chevalier94,chugai01, groh2014, dessart15,shivvers15,grafener16,yaron17,boian2019,boian2020}. The light curve will also be affected due to differences in the density structure of the outer regions and to the extra energy source from the interaction \citep{moriya2011,dessart17,forster19}. The variety in the CSM densities, compositions, extensions, and geometries, paired with the range of possible explosion properties, makes interacting SNe a diverse group. Some events show interaction for only a short period of time after explosion (hours to days; e.g. \citealt{khazov16}, SN 2013fs - \citealt{yaron17}, SN 2013cu - \citealt{galyam14}), while for others the interaction can last years (e.g. SN 1988Z - \citealt{turatto93}). Recent improvements in observational techniques have led to an increased number of early-time spectroscopic observations of interacting SNe arising from a variety of progenitors and giving rising to diverse SN types at late times \citep{graham19}.

SN 2016bkv was a peculiar event that has shown early-time interaction with a CSM \citep{hosseinzadeh2018,nakaoka2018}. What makes this event unusual is its membership to a rare class of transients (5\% of all type II SNe) known as low-luminosity SNe (\citealt{pastorello2004, lisakov2017}). These events are characterized by relatively low bolometric luminosities ($10^{7}-10^{8}$~\lsun), low expansion velocities (a few $1000$ \kms), low $^{56}$Ni ejecta masses ($<0.01$ \msun) and low explosion energies when compared to the average type II SN \citep{pastorello2004,spiro14}. 

The progenitors of low-luminosity SNe are currently highly debated, with suggestions that they could be relatively low mass ($8-12$ \msun) red supergiants (RSG) that are not massive enough to undergo core collapse and instead explode via electron capture \citep{chugai2000}. \citet{lisakov2018} considered all known low-luminosity SN- IIP to date and concluded that low-mass RSG progenitors are the most probable candidates, as high mass progenitors did not reproduce the length of the plateau. Pre-explosion images of several low-luminosity SNe have also suggested low-mass RSG progenitors based on their luminosities and effective temperatures (e.g. SN~2005cs - \citealt{maund05}, SN~2008bk - \citealt{vandyk12}, SN~2009md - \citealt{fraser11}).\citet{hira21} recently suggested that the progenitor of the low-luminosity event SN~2018zd was a super asymptotic giant branch (AGB) star that exploded via electron capture.

However, \citet{farrell20a} found that the final mass of a RSG is very uncertain based on its luminosity and effective temperature alone, allowing the possibility of more massive progenitors that evolved through binary interaction. An alternative suggestion is that the progenitors of low-luminosity SNe are high mass ($> 25$ \msun) RSGs ending in core collapse with significant fall back of material onto the newly formed black hole \citep{turatto1998, zampieri2003LLSN, utrobin2007}. \citet{turatto1998} modelled the light curve and spectrum of the low-luminosity SN 1997D and determined a progenitor mass of $25-40 M_{\odot}$.

Because it shows signs of interaction with material expelled by its progenitor, SN 2016bkv offers a great opportunity to constrain the properties of its progenitor star. The comparison with evolutionary models is also facilitated since we know the evolutionary stage (i.e. the star is at the end of its life). SN~2016bkv was discovered on 2016 March 21 in the galaxy NGC 3184, at a redshift of $z = 0.001975 \pm 0.000003$ (luminosity distance $d_{L} = 14.4 \pm 0.3$ Mpc; \citealt{itagaki16, hosseinzadeh2018}). The metallicity of the host galaxy is estimated at an average of $12 + \log (\mathrm{O/H}) = 9.07 \pm 0.12$ \citep{moustaka2010}. By fitting the initial photometric points, \cite{hosseinzadeh2018} estimates the explosion time ($t_{\mathrm{exp}}$) to be $1.2$ d before discovery. We adopt this estimate and throughout this paper we present all epochs with respect to $t_{\mathrm{exp}}$.

The lightcurve of SN 2016bkv peaks at $M_{\mathrm{V}} = -16$ mag, earning it the low-luminosity denomination. After its peak, it exhibits a long plateau phase ($\sim 140$ d; \citealt{nakaoka2018}) and a short drop to the radioactive tail. Measuring the luminosity during the radioactive decay phase, \cite{hosseinzadeh2018} find an unusually high $^{56}$Ni mass for this class of transients ($M_{\mathrm{Ni}} = 0.02$ \msun), while \cite{nakaoka2018} suggest a more typical $M_{\mathrm{Ni}} \approx 0.01$ \msun.  

The first spectra of SN 2016bkv were obtained $3$ days after the estimated post-explosion date, and for the first $\approx 5-8$ d post-explosion they show narrow emission lines characteristic of CSM interaction. After that, SN 2016bkv follows a typical low-luminosity SN photospheric and nebular phases. Its ejecta velocity is at most $2000$ \kms \citep{nakaoka2018,hosseinzadeh2018}, and was estimated from \ion{H}{$\alpha$} absorption profiles during the photospheric phase.

The properties of the progenitor of SN 2016bkv are not entirely clear. \citet{hosseinzadeh2018} compared the nebular spectra of SN 2016bkv to modelled spectra and concluded that the event was most likely the result of an electron capture SN from a low mass progenitor ($\approx 9$ \msun). On the other hand, using analytical models of typical type II-P SN light curves, \cite{nakaoka2018} derive an ejecta mass of $16.1-19.4$ \msun\ and a progenitor radius of 180--1080~\rsun. Based on the early-time luminosity, they also obtain a mass-loss rate of $\mdot = 1.7 \times 10^{-2}$ \msunyr and a CSM mass of $M_{\mathrm{CSM}} = 6.8 \times 10^{-2}$ \msun, assuming a wind velocity of $10$ \kms, and an ejecta velocity of $2000$ \kms. 
 
The goal of this paper is to obtain the mass-loss rate and surface abundances of the progenitor of SN 2016bkv by modelling its early-time spectrum. This will allow us to investigate the nature of the progenitor. We analyse the $4.1$ d spectrum of SN 2016bkv obtained with the Beijing Faint Object Spectrograph and Camera (BFOSC) mounted on the 2.16 m telescope at Xin-Glong. We download this spectrum via {\small WISEREP}\footnote{https://wiserep.weizmann.ac.il/} \citep{yaron12}. 

The paper is organized as follows. In Section 2 we discuss CMFGEN, the code we employ to model the observed spectrum of SN 2016bkv. We present the derived properties of the SN 2016bkv's progenitor in Sect. 3. In Section 4 we discuss the morphology of early-time low-luminosity SNe and perform a parametric study of the effects of mass-loss rate, luminosity and abundances. In Section 5 we discuss the presence of fluorescence CIII emission in the optical spectrum, while Section 6 explores the implications of our results on the nature of the progenitor of SN 2016bkv. Our concluding remarks are presented in Section 7.

%_________________________________________________________________

\section{CMFGEN Radiative Transfer Modelling}
\label{sect:methods}

To model the observed spectrum of SN 2016bkv we use the radiative transfer code {\small CMFGEN} \citep{hillier1998}. The code iteratively solves the coupled equations of radiative transfer and statistical equilibrium in non-local thermodynamic equilibrium, assuming spherical symmetry and stationarity. {\small CMFGEN} computes the full spectrum, including continuum and line formation. To convert the output of {\small CMFGEN} to the observers' frame, we use {\small CMF\_FLUX} \citep{busche05}. For further details on the implementation of {\small CMFGEN} in modelling interacting SNe and comparing with observations, we refer the reader to \citet{groh2014, shivvers2015, boian2019, boian2020}. 

The code takes a number of input variables, the most relevant in our models being the luminosity of the SN, $\lsn$, the mass-loss rate of the progenitor, \mdot, the terminal wind velocity of the progenitor, \vinf, and the abundances of each  included species throughout the CSM. If the photosphere is located in the CSM, the abundances that we derive should reflect those from the progenitor's surface shortly before explosion, as long as the CSM mass is relatively low ($<< 1~\msun$). This should be a good approximation specially if the progenitor was a RSG that has a convective envelope. We also specify the inner boundary, \rin, which represents how far into the CSM the SN ejecta has travelled at a given time $t$ after explosion. The region between the inner and outer boundary is divided up into 80 shells and {\small CMFGEN} calculates the required physical quantities, such as those needed for the level populations and the radiation field, at each shell.

We constrain the properties of SN 2016bkv by computing a new grid of models that span a range of \lsn, \mdot, and surface abundances for H, He, C, N, and O, and comparing the synthetic spectra to the observations. We include other chemical elements in our modelling, such as Si,  S, and Fe, with assumed solar abundances since their corresponding lines are not present in the observed optical spectrum. The other input parameters were chosen based on a number observed properties. The inner boundary is chosen so the modelled spectrum corresponds to a similar time after explosion as the observations. We assume the SN ejecta velocity was $2000 ~\kms$ \citep{nakaoka2018}. We note that this value is only an estimate since we assume the ejecta velocity is constant, when in fact it is expected to be decelerated by the interaction. In addition to the physical parameters described above, the density structure of the CSM $\rho(r)$ is derived from the mass conservation equation, and follows $\rho \propto r^{-2}$. At the inner boundary we assume a steep density gradient with a scale height of $h = 0.08\,\rin$ to mimic the SN shock front. The terminal wind velocity is set arbitrarily at $150~\kms$. Usually the width of the observed lines is set by \vinf\ however, the spectrum analysed here is limited by its low resolution of $v = 800$ \kms (resolving power $R \approx 375$; \citealt{hosseinzadeh2018}). For comparison with the observations, we downgrade the spectral resolution of our CMFGEN spectra by convolving with a Gaussian function with full-width half-maximum of $FWHM = 800$ \kms.

When determining the best-fitting parameters, we first compare the continuum normalized model spectra to the normalized observed optical spectrum to constrain the temperature at the base of the CSM (\tstar), \mdot, and the surface abundances. As discussed in \citet{boian2019} and \citet{boian2020},  \tstar\ affects the ionization structure of the CSM, \mdot\ the strengths of the lines, with the abundances affecting the strength of the emission lines of the relevant species. Then we compare the absolute flux to determine \lsn, where we assume the distance to SN 2016bkv to be $14.4 \pm 0.3$~Mpc \citep{hosseinzadeh2018}. We account for the interstellar extinction using the reddening law from \cite{fitzpatrick99}, and thus also constrain the best-fitting extinction parameter, $E(B-V)$, assuming a ratio of total-to-selective reddening of $R_{\mathrm{V}} = 3.1$.  To better understand the spectral morphology of low-luminosity interacting SNe, we also perform a parametric analysis to assess the effects of \lsn, \mdot, and chemical abundance on the spectrum. This also allows us to estimate their uncertainties, which we quote as estimated 90\% confidence ranges.

%______________________________________________________________

%__________________________________________________ One column table
   \begin{table}
    \centering
      \caption[]{The properties of the best fitting CMFGEN models to the observed $4.1$ d spectrum of SN 2016bkv.}
         \label{best_fit_parameters}
     $$ 
         \begin{array}{p{0.5\linewidth}l}
            \hline
            \noalign{\smallskip}
            Parameters &  \mathrm{Best-fit\,\,Value}   \\
            \noalign{\smallskip}
            \hline
            \noalign{\smallskip}
            SN Luminosity & (5.5 \pm 0.5) \times 10^{8} \ L_{\odot}   \\
            Temperature at base of CSM & 26100 ~\mathrm{K} \\ 
            Inner Boundary Radius  & (8.0 \pm 2.0) \times 10^{13} \ \mathrm{cm}\\
            Progenitor Mass-loss rate (for $\vinf=150~\kms$) & (6.0 \pm 2.0) \times 10^{-4} \ M_{\odot} yr^{-1}   \\
            Progenitor Wind Velocity &  \lesssim 800~\kms \\
            E(B-V) & 0.01~\mathrm{mag} \\
                        \noalign{\smallskip}
            \hline
            Species & \mathrm{Mass\,\,Fraction} \\
            \noalign{\smallskip}
            \hline
            Helium &   0.28 \pm 0.05   \\
            Carbon   &  0.003 \pm 0.001 \\
            Nitrogen    &  0.0011 \pm 0.0007 \\
            Oxygen  &  0.0010 \pm 0.0003 \\
            \noalign{\smallskip}
            \hline
         \end{array}
     $$ 
   \end{table}
%

%______________________________________________________________

\section{Properties of the Progenitor Star}
\label{sect:bestfit}

\begin{figure}
   \centering
   \includegraphics[width=1.07\linewidth]{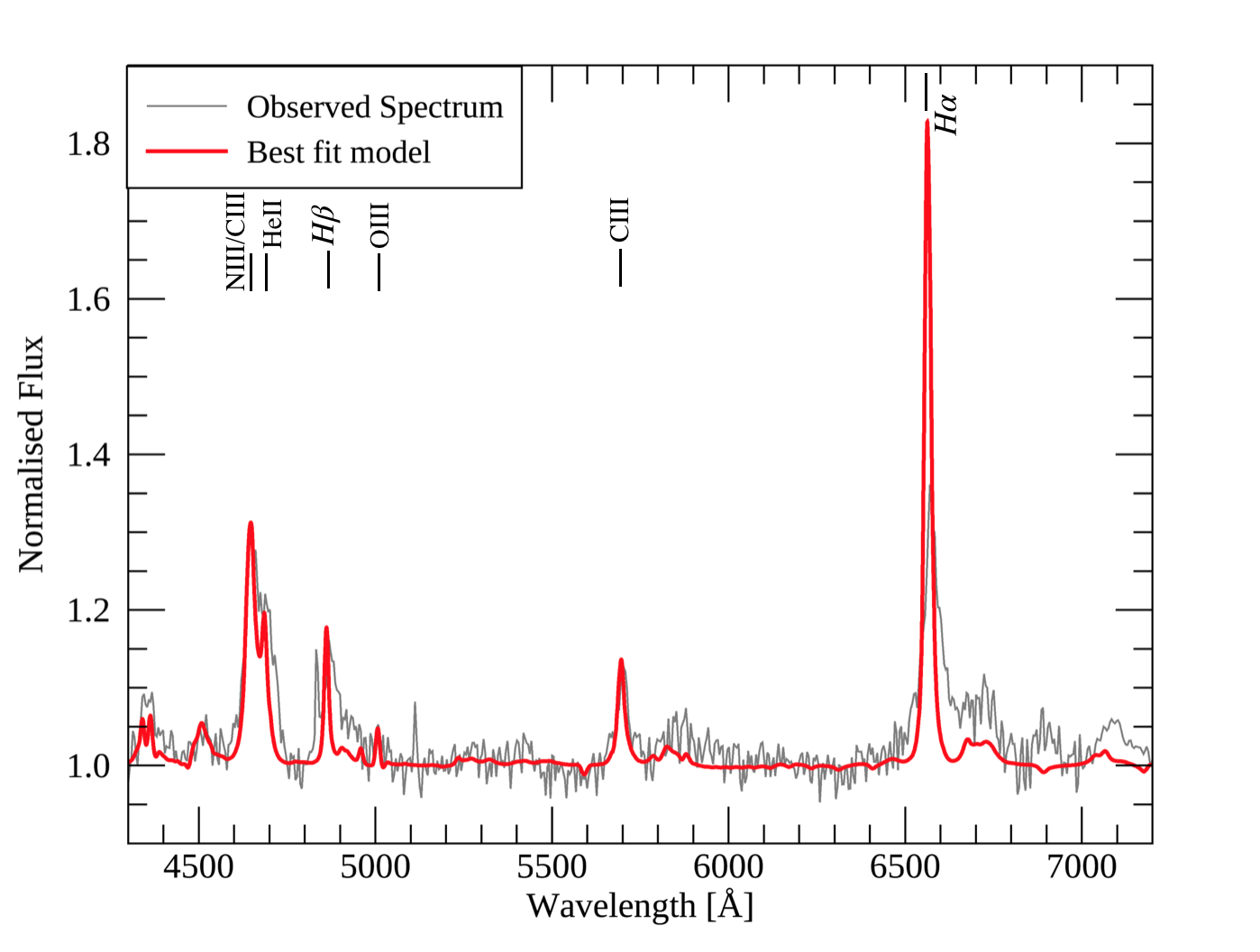}
      \caption{The spectrum of SN 2016bkv (grey) observed $4.1$ d post-explosion compared to our best-fitting CMFGEN model (red). The spectra have both been normalized to the continuum, and the strongest lines have been labeled.  
              }
         \label{bestfit}
   \end{figure}

Our models reproduce reasonably well the early spectrum of SN~2016bkv. Figure \ref{bestfit} shows the best-fitting model of SN~2016bkv compared with the observations, and our derived progenitor properties are listed in Table \ref{best_fit_parameters}. In this parameter range, small changes of the order of 30\% in the input parameters, such as \mdot\ or \lsn\, lead to noticeable changes in the strength of the spectral lines.

\subsection{Ionization of the progenitor's wind shortly after explosion}
\begin{figure}
\centering
    \includegraphics[width=1.0\columnwidth]{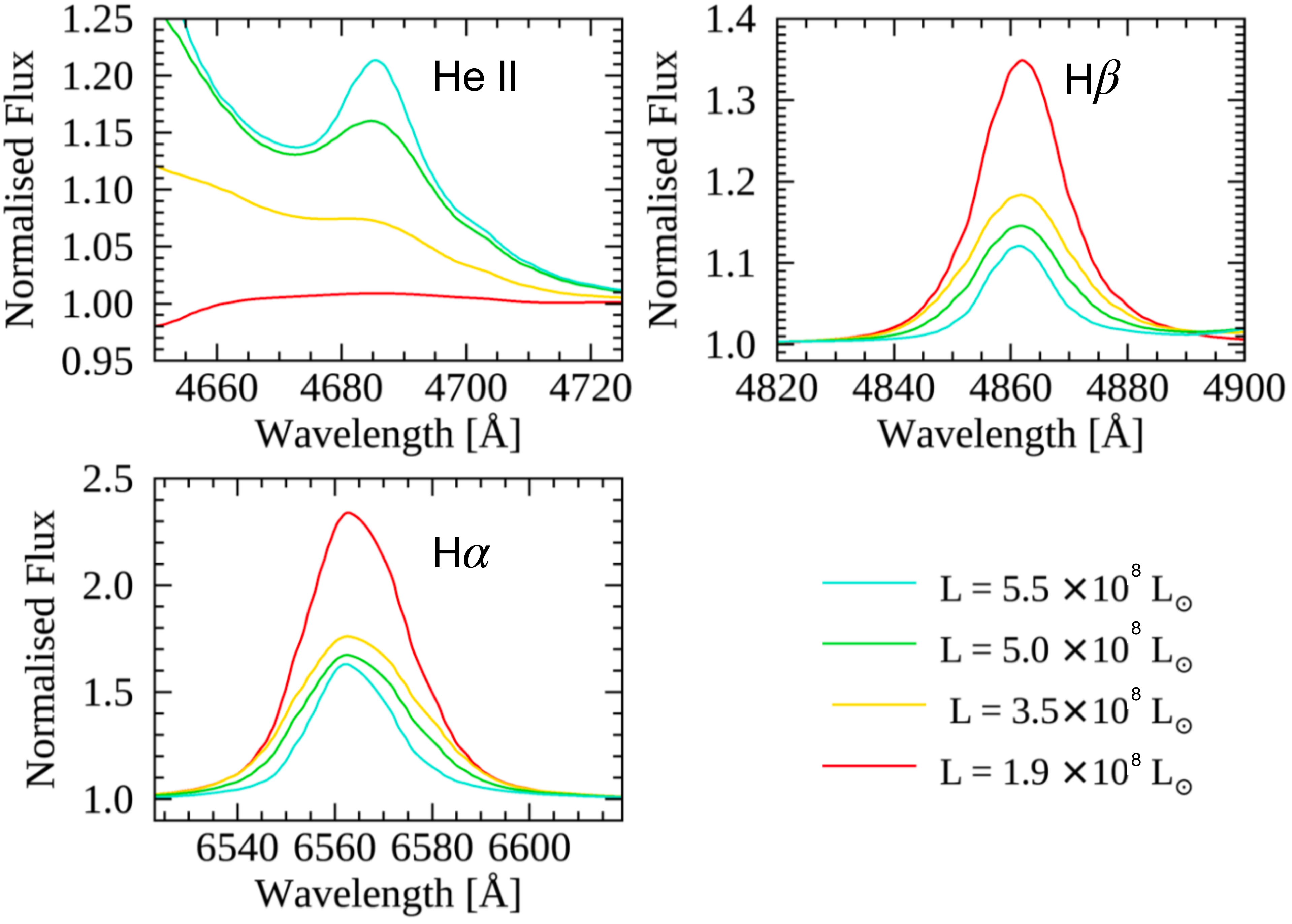}
    \caption{CMFGEN models exploring the effects of different SN luminosities on the spectra, while keeping all other parameters constant between the models. The different panels show \ion{He}{ii}~$\lambda4686$, H$\beta$,  and H$\alpha$. }
    \label{change_lum}
\end{figure}

Our analysis shows that \object\ had a luminosity of $\lsn= 5.5 \pm 0.5 \times 10^{8}$ \lsun\ at the time the earliest observed spectrum was taken ($t_{\mathrm{exp}} = 4.1$ d), and a  temperature at the base of the wind of $\tstar = 26100$ K. These values confirm its low-luminosity nature. The main diagnostic we used to obtain \tstar, and therefore \lsn\ for SN 2016bkv, is the ratio of \ion{C}{iii} to \ion{C}{iv} lines at 5696 and 5801~\AA. Changes in the ionisation structure are dictated by changes in the temperature structure, which are predominantly affected by \lsn\ \citep{boian2019}. Increasing \lsn\ (and thus \tstar, for a constant \rin) leads to a more ionized structure, i.e. decreasing the ratio of \ion{C}{iii} to \ion{C}{iv} emission lines, and increasing the strength of \ion{He}{ii}~$\lambda 4686$. We also find that increasing \lsn\ leads to a decreased strength of the \ion{H}{i} lines (Fig.~\ref{change_lum}). This happens because the effective recombination rate coefficient is inversely proportional to the temperature, $\alpha (H) \propto T^{-1/2}$ in this regime \citep{rybicki1979,boian2019}.

By comparing the synthetic CMFGEN flux with the flux-calibrated observed spectrum, we were also able to constrain the colour excess to $E(B-V) = 0.01$~mag, assuming a $R_{\mathrm{V}} = 3.1$, matching well the upper limit of $E(B-V) \leq 0.015$~mag determined by \cite{nakaoka2018} using the \ion{Na}{i} line. 

We also explore the effects of the inner radius, \rin\ on the emergent spectra. Increasing \rin\ leads to a weakening of most emission lines, as \rin\ is inversely proportional to the density (from the mass continuity relation). In practical terms, we can think of this effect as moving further out from the centre of the explosion, i.e. to lower CSM densities.

\subsection{Mass-loss rate in the last few years before explosion}
Our interpretation is that the CSM around \object\ is composed of material ejected by the progenitor star before the explosion, in agreement with the scenario proposed by \citep{hosseinzadeh2018}. Our best-fitting model points to the progenitor of SN~2016bkv having had a mass-loss rate of $\mdot = 6.0 \pm 2.0 \times 10^{-4}$ \msunyr\ in the few years prior to the explosion, assuming $\vinf = 150$ \kms. This value of \mdot\ is just above the lower limit for observable CSM signatures \citep{boian2019}, which explain the weakness of the observed emission lines. Due to the low resolution of the spectrum, we cannot fully constrain \vinf\ and \mdot, but only place an upper limit of $\vinf < 800$ \kms and $\mdot < 4.3 \times 10^{-3}$ \msunyr.To be able to compare to other events, we can use instead the wind density parameter, $D = \frac{\mdot}{4 \pi \vinf}$ \citep{chevalier11}. For SN~2016bkv we find $D = (1-2) \times 10^{14}~ \mathrm{g\, cm^{-1}}$, which is well below the average value of  $D \approx 5 \times 10^{16}~ \mathrm{g\, cm^{-1}}$ for `normal' luminosity interacting SNe \citep{boian2020}. Our mass-loss rate determined from fitting the spectrum showing SN-CSM interaction signatures is significantly lower than the previously estimated value from \cite{nakaoka2018} of $\mdot = 1.7 \times 10^{-2}$ \msunyr (for an assumed $\vinf  = 10$ \kms). However the latter is based on the luminosity-mass loss rate analytical relation for interacting SNe, which contain strong assumptions regarding the exact luminosity contribution from the interaction, and the energy conversion efficiency. 

We use \ion{H}{$\beta$} as the main diagnostic for \mdot.  It should be noted that the \ion{H}{$\beta$} line has an apparently double peaked profile in the observations, a morphology that is not matched by the models. This could be an indication of a disk or torus-like structure which could be the result of binary interaction. However, since the blueshifted peak is not seen in the spectrum taken 4.8~d after explosion \citep{hosseinzadeh2018} or in other Balmer lines, it is more likely that the blueshifted peak is an artefact in the observations. 

While our best-fitting model with $\mdot=6.0 \times 10^{-4}$ \msunyr\ reproduces well most of the lines observed in SN 2016bkv, however it significantly overestimates the peak of the \ion{H}{$\alpha$} emission line while underestimating the wings (Fig. \ref{bestfit}). To fit the equivalent width of \ion{H}{$\alpha$} we would need a slightly higher value of $\mdot\simeq8.0 \times 10^{-4}$ \msunyr, which is still lower than the value quoted by \citet{nakaoka2018}. 

\begin{figure}
   \centering
   \includegraphics[width=1.0\columnwidth]{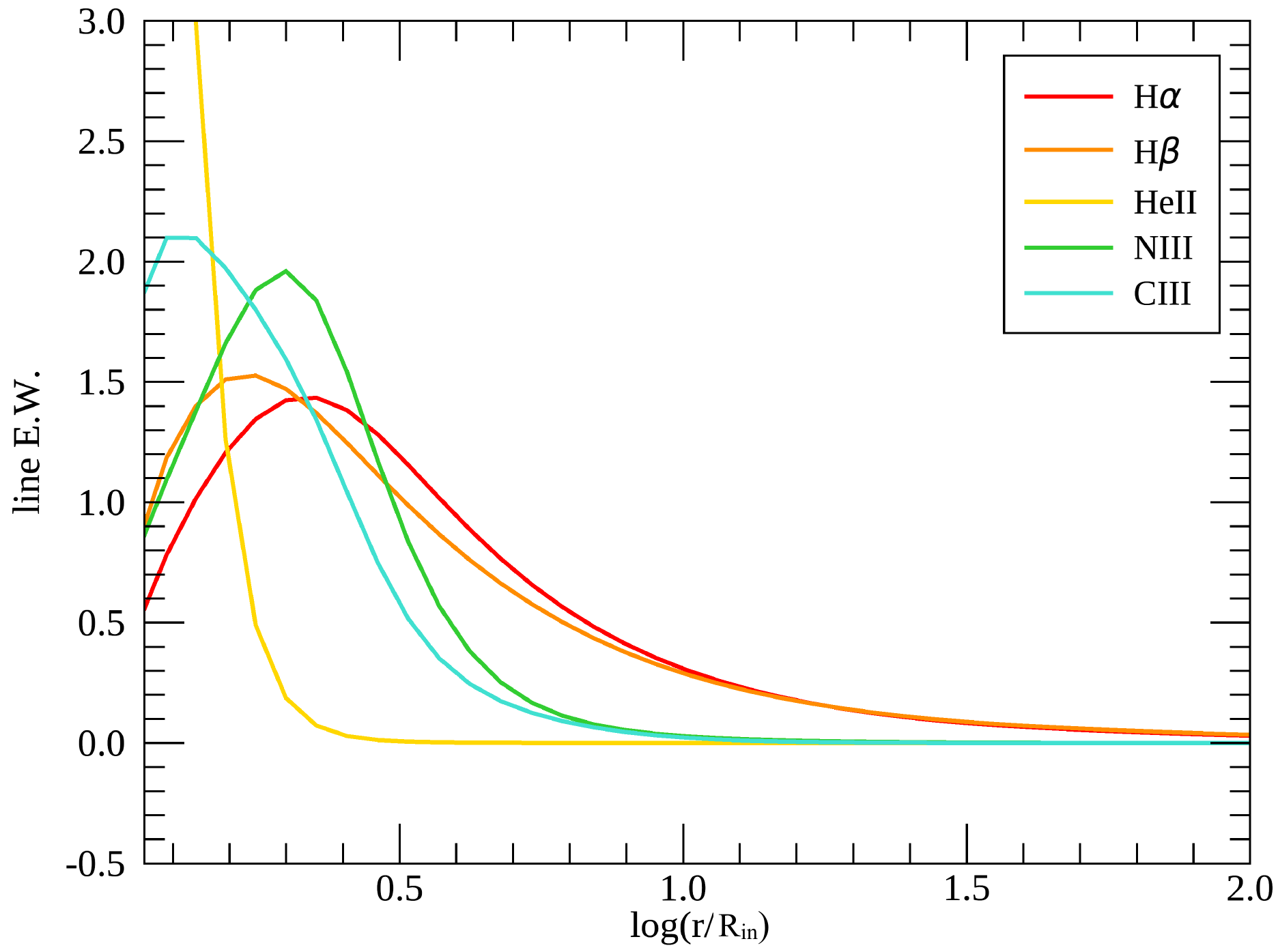}
      \caption{The formation regions for the most prominent emission lines in the spectrum of SN 2016bkv. The x-axis corresponds the distance $r$ from the center of the explosion, normalized to the inner boundary radius. The y-axis is proportional to the equivalent width of a given line, and is given in arbitrary units and normalized.}
         \label{line_formation}
   \end{figure}

A discrepancy between \ion{H}{$\alpha$} and \ion{H}{$\beta$} is also seen in SN~1998S \citep{shivvers15}, where it was interpreted as due to time-dependent effects on the H level populations. It possible that the same happens for \object. Alternatively it is also possible that the density structure is steeper than we assume in our models, or follows a non-spherical mass distribution. If the CSM were truncated, this would likely affect \ion{H}{$\alpha$} and \ion{H}{$\beta$} almost equivalently as we can infer from their line-formations regions (Fig.~\ref{line_formation}).

Based on our CMFGEN results, we can also calculate lower limits for the CSM extension, mass, and duration of the mass-loss episode. We find the lower limit of the CSM extension by locating the largest distance at which a line is formed in our best-fitting model (Figure \ref{line_formation}). Commonly \ion{H}{$\alpha$} is used for this purpose but as we discussed above, this line is affected by time dependent effects and not well matched by our model. Instead we use the \ion{C}{iii}$\lambda 5696$ line which is well fitted by our models. We find $r_{\mathrm{CSM}} \geq 10\  \rin$, where $\rin = 8.0 \times 10^{13}$ cm. Assuming $\vinf = 150$ \kms\ we find a mass-loss duration of $\geq 1.7$ yrs. Since $\mdot = 5.0 \times 10^{-4}  M_{\odot} \ yr^{-1}$ ($v_{\infty}= 150$ \kms), we find a CSM mass of $M_{\mathrm{CSM}} \geq 0.00085$ \msun.

\begin{figure*}
\centering
    \includegraphics[width=0.75\textwidth]{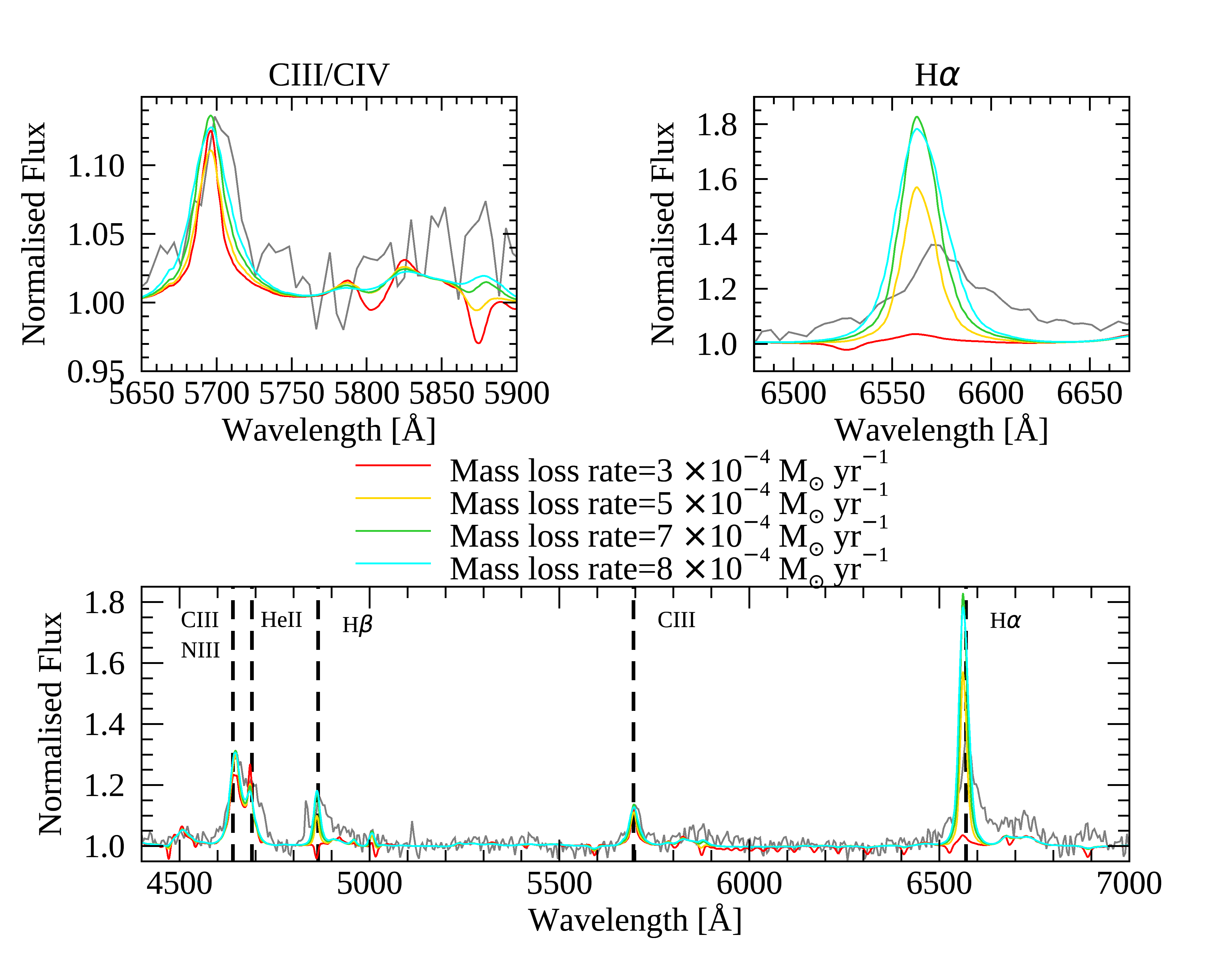}
    \caption{CMFGEN models exploring the effects of different progenitor mass-loss rates on the post-explosion spectra, while keeping all other parameters constant between the models. The observed $4.1$ d spectrum of SN 2016bkv is also shown (grey). Top left: zoom-in around the \ion{C}{iii} and \ion{C}{iv} lines. Top right: zoom-in around the \ion{H}{$\alpha$} region. Bottom: full optical range.}
    \label{change_mdot}
\end{figure*}

To understand how \mdot\ impacts the spectral morphology, we plot in Fig.~\ref{change_mdot} CMFGEN models computed for different values of \mdot\ while keeping the other parameters the same as in the best fit model for \object.  We find that increasing \mdot\ leads to stronger emission lines as it increases the density of the CSM. However, if we increase \mdot\ past $7.0 \times 10^{-4}$ \msunyr\ we see a decrease in the strength of \ion{H}{$\alpha$}. This is because when there is a large amount of hydrogen recombination, neutral hydrogen begins to dominate, reducing hydrogen emission \citep{boian2019}. Increasing \mdot\ while keeping the influx of photons at the inner boundary constant also results in a less ionized structure since the temperature in the outer CSM will be reduced.

\subsection{Surface abundances of the progenitor}

\begin{figure}
    \centering
\includegraphics[width=0.99\columnwidth]{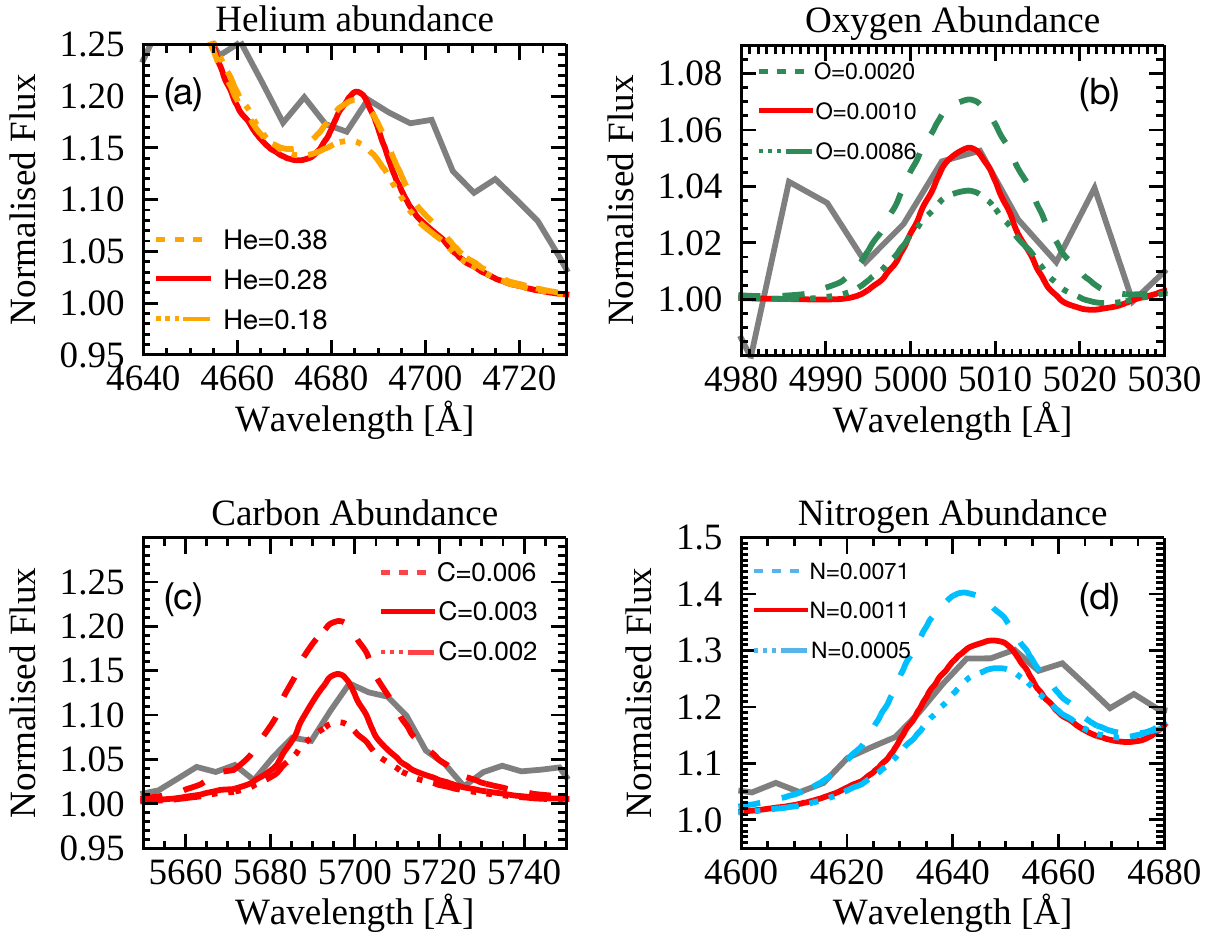}
    \caption{CMFGEN models exploring the effects of varying the abundance of one species at a time as indicated by the labels.  Each panel shows the observed $4.1$ d spectrum of \object\ (grey), the best-fit model (red), and additional models with the same parameters as the best fit model except for the abundance of the species indicated by the legend  (in mass fraction). We present close-up regions around the following relevant lines: \ion{He}{ii} (top left), \ion{O}{iii} (top right), \ion{C}{iii} (bottom left), and \ion{N}{iii} (bottom right).
    \label{chemical_abun}}
\end{figure}

Next we investigate the effect of varying the abundances of He, C, N, and O, since these species have strong spectral lines in the optical. We proceed systematically by producing CMFGEN models in which we separately vary the abundance of one element at a time, keeping the other parameters fixed using the values from the best-fit model quoted in Table~\ref{best_fit_parameters}. This reference best-fit model is shown in red in all panels of Fig. \ref{chemical_abun}. For each element, as we increase its abundance, the strengths of its corresponding emission lines increase as expected (Figure \ref{chemical_abun}). 

Below we discuss the diagnostics and individual abundance determinations. Within the errors, we find that the He, C, N, and O abundances match the solar values from \citet{asplund2005}. Our results are somewhat surprising since typically evolved massive stars that explode as core-collapse SNe have CNO-processed material at the surface, i.e. high N abundance \citep{groh2013,boian2020}. In addition to \object, out of a sample of 21 interacting SNe that have been analyzed in detail with CMFGEN, only 2 other interacting SNe have progenitors with solar surface abundances at the pre-SN stage, namely SN~2013fs \citep{yaron17} and SN 2018zd \citep{boian2020}.

Figure~\ref{chemical_abun}a shows the effect of different He abundances in the \ion{He}{ii}~$\lambda 4686$ line. Combined with the constraints for fitting \ion{H}{$\beta$}, we estimate the He mass fraction as $0.28\pm0.05$. We find that the most sensitive diagnostic of the oxygen abundance is the \ion{O}{iii}$\lambda 5007$ spectral line. We constrain the oxygen mass fraction to $0.0010\pm0.0003$ (Fig.~\ref{chemical_abun}b).  We use the strongest N line in the optical spectrum (\ion{N}{iii}$\lambda 4640$) to estimate a N mass fraction of $0.0011\pm0.0007$ (Fig.~\ref{chemical_abun}d). We note however that this region is contaminated by \ion{C}{iii}$\lambda4650$ emission. Still, even considering this systematic uncertainty as discussed below, our models rule out N mass fractions higher than 0.002. 

We use the isolated \ion{C}{iii} $\lambda 5696$ line to constrain the C mass fraction to $0.003\pm0.001$ (Fig.~\ref{chemical_abun}d). This estimated C abundance also provides consistent results for the blended \ion{C}{iii} $\lambda 4650$ (Fig.~\ref{chemical_abun}c), giving further weight to our C abundance determination. As shown in Fig.~\ref{chemical_abun}c, the strength of \ion{C}{iii} $\lambda 5696$ is extremely sensitive on the C abundance, and the statistical error on fitting the line equivalent width is much lower than the 0.001~dex quoted above. 

We adopt a conservative error of 0.001 dex in the C abundance to account for the systematics arising from non-LTE radiative transfer effects. The \ion{C}{iii} $\lambda 5696$ line is affected by C and S lines in the UV \citep{martins2012}. For this reason, it is crucial that we included S in the model atom to mitigate the systematic effects on the C level populations. The magnitude of the effects found in \citet{martins2012} are applicable to the outflows of O-type stars, and detailed, extensive systematic tests in a similar fashion would be needed to precisely quantify the effects for interacting SNe conditions. We investigate in Sect.~\ref{LINES} the formation of these optical C lines in our physical situation, and it is reassuring that the \ion{C}{iii} $\lambda 5696$ line is mainly formed by recombination. Based on our models with different \lsn\ (and thus different far-UV fluxes), we estimate that our error of 0.001 dex in the C abundance accounts for the systematics when using the \ion{C}{iii} $\lambda 5696$ line as we did. In any case, this systematic uncertainty still means that if we are overestimating (underestimating) the C abundance, we may be underestimating (overestimating) the N abundance using the line at $4640$ \AA. For this reason, to be conservative we estimated a relatively high relative uncertainty in the N abundance above. However, even though the \ion{C}{iii}$\lambda 4650$ and \ion{N}{iii}$\lambda 4640$ lines cannot be disentangled due to the resolution of the spectrum, because the \ion{C}{iii}$\lambda 4650$  line lies at a slightly redder wavelength compared to the \ion{N}{iii}$\lambda 4640$ line, if the C abundance would have been severely misjudged, the red side of the peak would not fit the observations (Fig.~\ref{chemical_abun}c,d).

\section{Fluorescence CIII emission in early-time supernovae} \label{LINES}    

Fluorescence lines occur when an atom is promoted to an excited state by colliding with a photon of the correct energy and later spontaneously decays. A recombination line occurs if an element is ionized and re-encounters an electron to recombine and emit a photon. Therefore the strength of a fluorescence line depends on the density of photons and protons, while a recombination line depends on the density of protons and electrons. This makes fluorescent lines respond more strongly to changes in the radiation field, i.e. in \lsn, and recombination lines to changes in the gas density.

We noticed that the \ion{C}{iii} lines at $4650$ \AA\ and at $5696$ \AA\ respond to changes in the wind properties differently, indicating that they are not formed through the same mechanism. We found that in the physical conditions of interacting SNe, the \ion{C}{iii}$\lambda 4650$ line responds more strongly to changes in \lsn, while the \ion{C}{iii}$\lambda5696$ line is more affected by changes in density, which are in turn controlled by \mdot, \rin, and the C abundance. We propose that the \ion{C}{iii}$\lambda 4650$ line is formed through a fluorescence process, while the \ion{C}{iii}$\lambda5696$ emission is formed mainly due to recombination. 

\begin{figure}
   \centering
    \includegraphics[width=1.0\columnwidth]{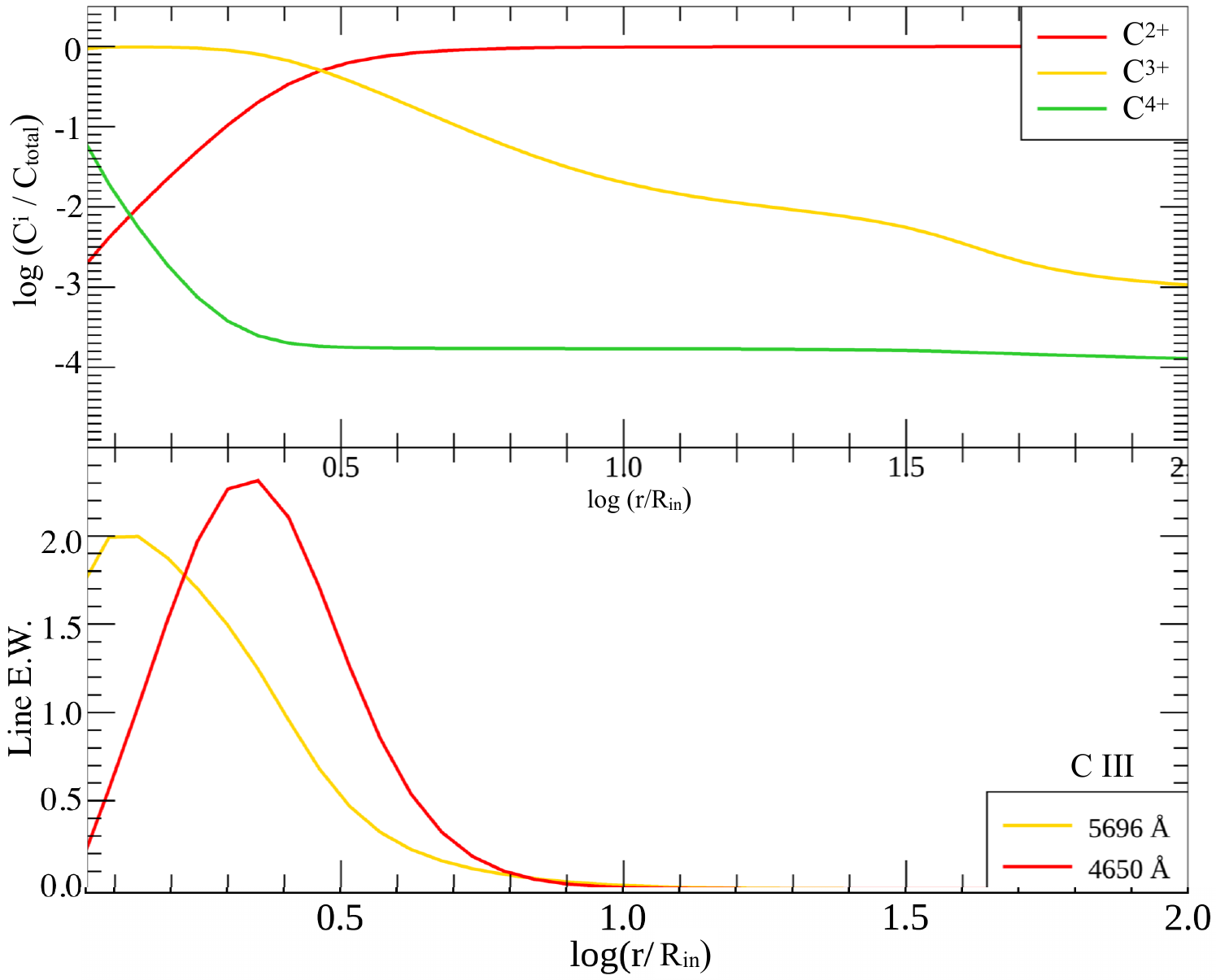}
      \caption{ Top panel: Number density of different C ions divided by the total number of C nuclei as a a function of distance to the center of the explosion (normalized to the inner boundary radius). Bottom panel: the line formation regions of the \ion{C}{iii}$\lambda 4650$ line (red) and the \ion{C}{iii}$\lambda 5696$ line (yellow). The y-axis is proportional to the normalized equivalent width of a given line. }
         \label{CIII_line_formation}
   \end{figure}

To investigate this further, we show the formation regions of the two \ion{C}{iii} lines in question, together with the  relative abundance of \ion{C}{} ions over those regions in Figure \ref{CIII_line_formation}. We find that in the region where the line at $4650$ \AA\ is formed, the C$^{2+}$ ions are competing with  C$^{3+}$ ions, whereas in the region where the line at $5696$ \AA\ is formed,  C$^{3+}$ ions dominate. This aligns with the proposition that the line at $5696$ \AA\ is a recombination line and the line at $4650$ \AA\ is a fluorescence line. 

This is an important result to keep in mind when interpreting the spectral morphology of interacting SNe. Also, one could potentially derive an erroneous abundance for certain elements if one relies on a single diagnostic line, in particular \ion{C}{iii}$\lambda 4650$. This has implications for constraining the SN progenitor nature as we discuss next.

%----------------------------------------------------------- 
%______________________________________________________________

\section{Discussion: implications for the origins of low-luminosity SN~II-P}
\label{sect:disc}

The early spectrum of \object\ provides the first opportunity to probe the surface abundances of the progenitor of a low-luminosity SN. This allows us to constrain the progenitor nature by comparison with stellar evolution models. Our analysis of SN 2016bkv showed that its progenitor had solar-like surface abundances of He, C, N, and O at the pre-explosion stage. We discuss next the possibilities of producing this abundance pattern under single or binary star evolution. 

\subsection{Single-star evolution requires an odd red supergiant}
\label{sect:singleEvol}
\begin{figure}
   \centering
      \includegraphics[width=0.99\columnwidth]{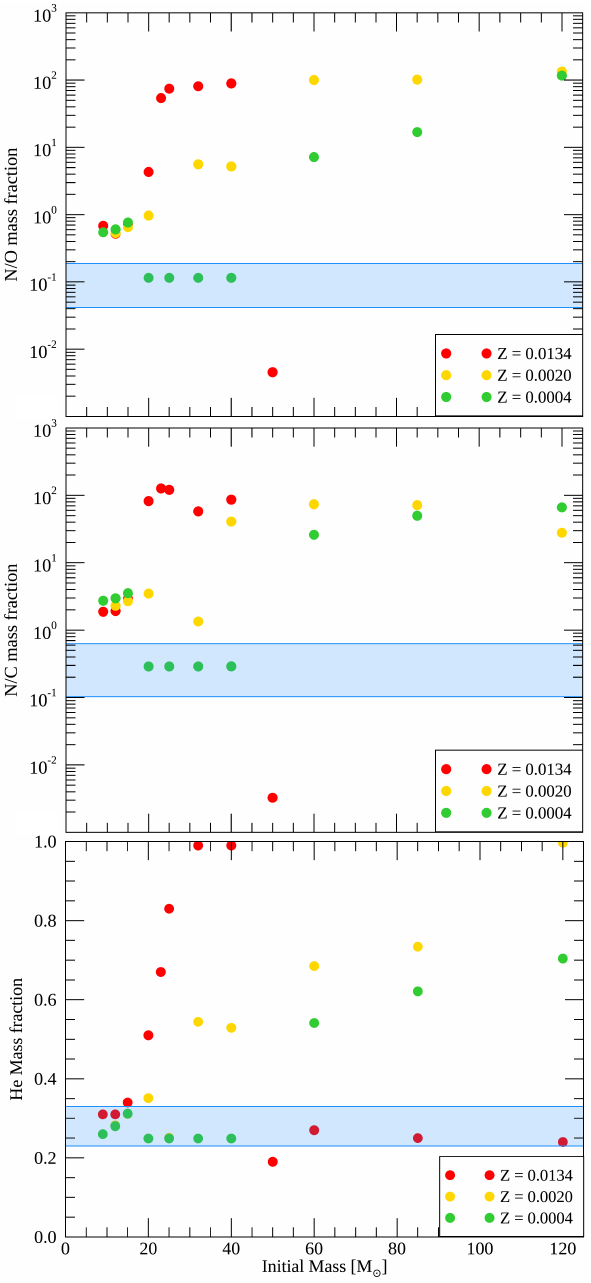}
      \caption{From top to bottom, surface abundance mass ratios of N/O and N/C, and He surface mass fraction at the pre-SN stage as a function of initial mass for Geneva stellar evolution models at 3 metallicities. The parameter space corresponding to the best-fit CMFGEN models of SN 2016bkv is highlighted in blue.
              }
         \label{N_O}
   \end{figure}

We compare the He, N, C and O surface abundances we found through CMFGEN modelling of SN 2016bkv to values from Geneva stellar evolution models of massive stars (Fig.~\ref{N_O}). We explore a large range of initial masses and three different metallicities, namely $Z = 0.014$ \citep{ekstrom2012}, $Z = 0.002$ \citep{georgy2013}, and $Z = 0.0004$ (\citealt{groh19}). None of the rotating models fit the abundance pattern that we found for the progenitor of \object, since they have too much N at the surface compared to \object. Thus, we focus the rest of the discussion on non-rotating models, that can also be interpreted as a proxy for models with very low rotational mixing.

Figure~\ref{N_O} shows that for solar and SMC metallicities the predicted values of N/O, N/C and He are in general too high compared to that inferred from the observations (blue band). There is a tiny transition region between the 40 and 50~\msun\ models at solar metallicity that would match the observation if a model were fine tuned. The progenitor in this region would be a transition between LBVs and WN stars \citep{groh2013}. However, the H envelope mass of this progenitor would be extremely low and likely incompatible with a SN~II-P such as \object. Thus we reject this possibility. 

Red supergiants evolving from slow rotators between 9--15~\msun\ at solar $Z$ would match our derived value of the surface He abundance and retain a large H envelope before exploding, which would produce a SN~II-P \citep{dessart09,goldberg19}. However, the N/C and N/O abundances are too high in the models (Fig.~\ref{N_O}). These are non-rotating models (thus no rotational mixing is present), and the enhancement in N/C and N/O is caused by dredge up. This happens when the bottom of the convective envelope of a RSG reaches the deep layers with CNO-processed material. Since the convective envelope is assumed to mix efficiently, signatures of CNO-processed material appear at the surface of these models, and this is inconsistent with our observations of \object. 

While the very-low metallicity Geneva models ($Z = 0.0004$; green points in Fig.~\ref{N_O}) match the N/O and N/C, they do not match the absolute values of C, N, and O, which are around solar. However, the host galaxy of SN 2016bkv (NGC 3184; \citealt{itagaki16}) is a spiral galaxy of the SABcd type and has an average oxygen content of $12 + \log \mathrm{(O/H)} = 9.07 \pm 0.12$ \citep{moustaka2010}, i.e. slightly higher than the solar value. Therefore we dismiss this scenario.

To explain the surface abundances of the progenitor of SN~2016bkv, it appears that modifications to the current understanding of convective mixing in stellar evolution models would be required if the progenitor was a red supergiant. We now speculate how single-star models and observations of \object\ could be reconciled. One possible way would be if the star did not become a RSG, and thus no deep convective layer extending to the surface would be present \citep{farrell20b,farrell20a}, and no dredge up of CNO-processed material would occur.  Another possibility is that the star was an odd RSG with no deep convective layer, and thus had no dredge up as well. Detailed models are obviously required to investigate if this scenario works or not. In addition, the models would need to match the other observational constraints available for massive stars such as the main-sequence width and evolution of rotational velocities.

\subsection{A progenitor that evolved through merger or binary mass transfer}
A large fraction of massive stars are found in multiple systems \citep{sana2012,moe2017}. Binary interactions can significantly change the evolution of a star and produce a diversity of SN types \citep{Podsia92,Podsia2017,eldridge18}. If SN 2016bkv is the result of a binary interaction, it would have been a different interaction from that of SN 1987A. SN 2016bkv was not enhanced in N (Sect. \ref{sect:bestfit}). Also, it had a long plateau phase and a fast rise to the maximum \citep{hosseinzadeh2018} whereas SN 1987A was N enhanced, had an almost non-existent plateau phase and a slow rise in the lightcurve \citep{arnett89,mattila10}.

We now discuss if the properties of the progenitor of \object\ can be explained under binary evolution. We acknowledge that detailed binary population synthesis would be needed to infer the likelihood of the different scenarios. Our conclusion that a RSG with a deep convective envelope would have enhanced N at the surface, and thus be inconsistent with the observations of \object, should still hold for stars that interacted in a binary system. If binary interaction resulted in the progenitor having a low core mass ratio, this would favor a blue supergiant solution \citep[e.g.][]{Podsia92}, given the relationship between the core mass ratio and the formation of a red supergiant  \citep[e.g.,][]{lauterborn71,maeder81, chiosi78, farrell20b}. The progenitor of \object\ does not necessarily need to be a BSG -- a yellow supergiant solution may also be consistent with our abundance pattern, as long as the star is hot enough to avoid dredge-up.

In a mass-transfer scenario, the progenitor would have to have accreted unprocessed material from a companion to reproduce our inferred low surface N abundance. Also the subsequent mass loss needs to be low so the material is retained at least in part until core collapse. A merger event can have a range of different outcomes depending at what stage of evolution the merger occurs \citep[e.g.][and references therein.]{zapartas19}. 

In the case of a post-Main Sequence + Main Sequence merger (case B), a strong molecular weight gradient could have been established before the merger event, which could prevent significant rotational mixing \citep{mestel53, mestel57, mestel86,justham2014}. This would in principle favor a low surface N abundance as observed in the progenitor of  \object. However, as pointed out to us by the referee, the merger models from \citet{justham2014} that avoid the RSG stage also seem to show some dredge-out of their post-MS helium cores as a consequence of the merger. Further work is needed to evaluate if this dredged-out material reaches the surface, either via mixing or post-merger mass loss. If the mixed material reaches the surface, the surface N abundance would increase.

\begin{figure}
   \centering
   \includegraphics[width=0.99\columnwidth]{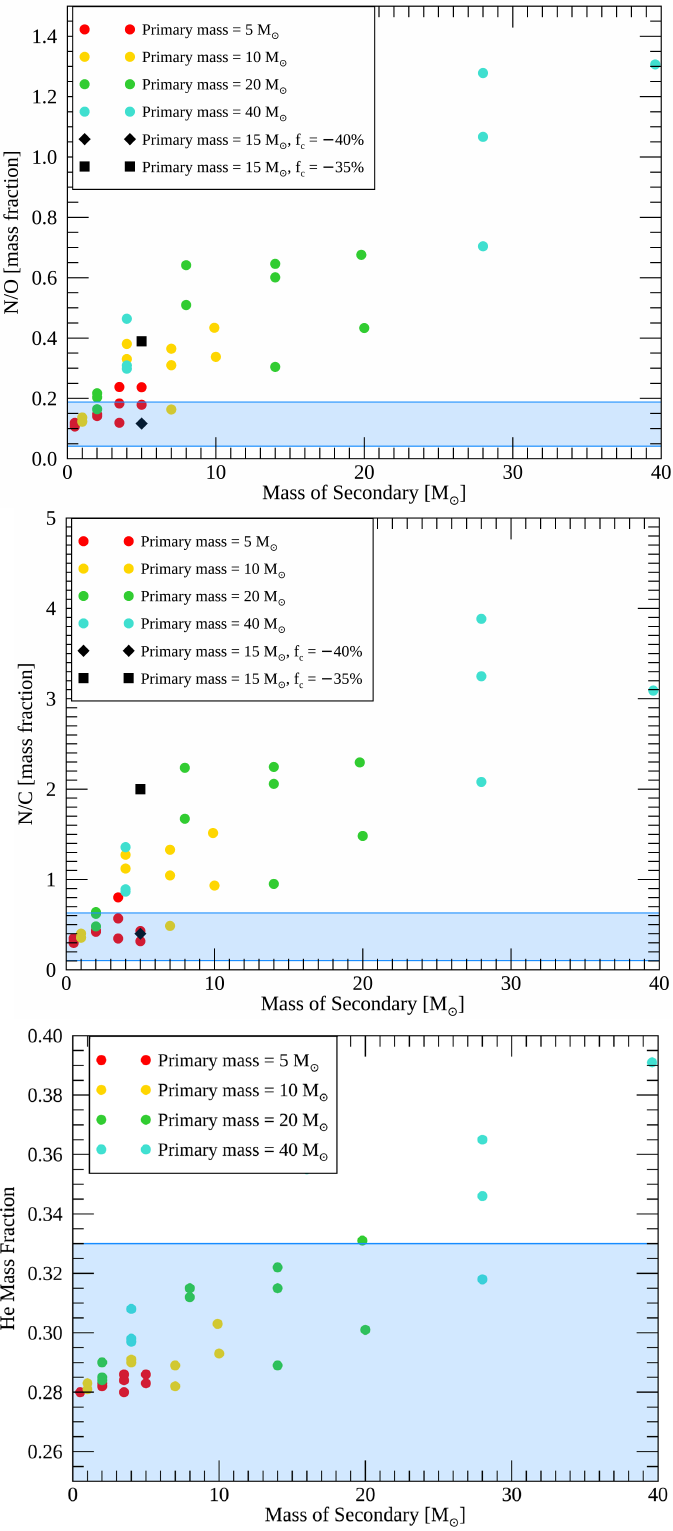}
      \caption{From top to bottom, surface abundance mass ratios of N/O and N/C, and He surface mass fraction as a function of secondary mass for stellar evolution models of mergers. The red, yellow, green and cyan correspond to the \citet{glebbeek2013} models, corresponding to merger product once equilibrium has been established at the beginning of He burning. The diamond and square black shapes represent data from \citet{menon2017b} and represent the pre-SN abundances. The $f_c$ values quoted describe the fraction of He core dredged up. Negative $f_c$ values imply that the He core of the primary has been penetrated.
              }
         \label{merg_N_O}
   \end{figure}

Keeping the merger scenario in mind, we compare in Fig.~\ref{merg_N_O}  the surface N/O and N/C ratios, and the surface He abundance of SN 2016bkv's progenitor to those from stellar evolution merger models from \citet{glebbeek2013}. Those models simulate the violent mergers of two stars at different points during the main sequence for a variety of primary and secondary masses. Those models predict the He, C, N and O abundances only for the immediate post-merger product once thermohaline mixing has modified the surface abundances. This is still earlier in the evolution (He burning) -- nonetheless, we can assume that the models provide lower limits to the surface N/C and N/O at the pre-SN stage. This is due to the fact that mass loss, rotational mixing and convection will tend to increase the amount of CNO-processed material at the surface, and thus N/C and N/O generally increase as a star with a H-rich envelope evolves. Therefore, we can rule out the merger models that do not fit the observed N/C and N/O.  However, it is possible that models that match the observations will no longer do so if they were evolved to their pre-SN stage. 

With these caveats in mind, we find four merger scenarios that have He cores massive enough to explode as SNe and simultaneously match the surface abundance that we found for \object's progenitor (Table \ref{mergers}). The models correspond to the merger of 10 + 7~\msun\ stars with the primary star halfway through the main sequence, 10 + 1~\msun\ stars with the primary either towards the terminal stage of the main sequence or core hydrogen exhaustion,  and 20 +  2~\msun\ stars with the primary star at core hydrogen exhaustion. The merger products consistent with \object's progenitor have a high He core mass compared to a single star of the same mass and evolve to become RSGs \citep{glebbeek2013}.  Given the relatively weak [\ion{O}{I}] that was observed in the nebular phase of \object\ \citep{hosseinzadeh2018}, in principle the models with the lowest He core masses, i.e. those with a 10~\msun\ primary star, would be favored. Further models would be needed to assess the final CNO abundance of these merger models, but they are possible candidates to explain the evolution of the progenitor of \object. 

There is also extensive work relating BSGs evolving through merger or mass transfer scenarios to SN~II explosions, with the classical example being SN~1987A \citep[e.g.][]{arnett89, Podsia92, menon2017b}. We compare our results to the models from \citet{menon2017b}, which were developed in the context of SN 1987A and provide the surface abundances at the pre-SN stage. The only model that falls within our parameter space has a negative $f_c$ value, implying that the He core of the primary is penetrated in the merger process \citep{menon2017b}.  This model has low N abundances at the pre-SN stage and with N/C and N/O ratios consistent with those that we found for \object. Our results provide an important comparison point to SN 1987A, and we encourage further massive merger models that could reproduce our abundance results simultaneously to the SN lightcurve and late-time nebular spectrum.

     \begin{table}
      \caption[]{Table showing the merger scenarios that fall in the parameter space of SN 2016bkv. $M_1$ is the primary mass, $M_2$ is the secondary mass, $M_\mathrm{merger}$ is the total mass of the merger product and $M_\mathrm{He}$ is the mass of the He core at He ignition. The evolution stage is that of the primary at the time of merger.}
         \label{mergers}

         \begin{tabular}{c  c  c  c  l}
            \hline
            \noalign{\smallskip}
            $M_1$ & $M_2$ & $M_\mathrm{merger}$ & $M_\mathrm{He}$ & Stage that merger occurs    \\
            (\msun) & (\msun) & (\msun) & (\msun) &\\
            \noalign{\smallskip}
            \hline
            \noalign{\smallskip}
            10.0 & 1.0 & 10.81 & 2.49 & Core Hydrogen Exhaustion \\
            10.0 & 1.0 & 10.83 & 2.43 & Terminal Age Main Sequence \\
            10.0 & 7.0 & 16.00 & 4.24  & Half Way Main Sequence \\
            20.0 & 2.0 & 21.04 & 6.49 & Core Hydrogen Exhaustion \\

            \noalign{\smallskip}
            \hline
         \end{tabular}

   \end{table}

\section{Concluding Remarks}

In this work we use the radiative transfer code CMFGEN to model the early-time spectrum of the interacting low-luminosity SN 2016bkv. We explore a large range of parameters in order to constrain the progenitor properties of SN 2016bk, and we summarize our main findings below. 

\begin{enumerate}

    \item The best-fitting CMFGEN model to the $4.1$ d spectrum of SN 2016bkv suggests that the progenitor had a relatively low mass-loss rate compared to other interacting events, of $\mdot = 6 \pm 2.0 \times 10^{-4}$ \msunyr, assuming a terminal wind velocity of $\vinf = 150$ \kms. 
    \item While the majority of the emission lines observed in the spectra of interacting supernovae are formed by recombination processes under radiative equiliibrium, and thus have a strong density dependence, other mechanisms such as fluorescence may power some lines, in which case these lines respond more strongly to luminosity variations. We find that is the case for the \ion{C}{iii}$\lambda 4650$ line in the early-time spectrum of SN 2016bkv. We thus recommend that multiple lines should be employed when estimating the C abundances of such events.
    \item The models also reveal that prior to explosion the surface abundances of SN 2016bkv's progenitor are consistent with solar contents of He, C, N and O. This contrasts with the CNO-processed material that is often found in progenitors of interacting SNe \citep{boian2020}.
    \item We compared our surface abundance results for \object\ to those predicted by single and binary star models.  The current single-star models do not reproduce our results. If the progenitor of \object\ evolved as a single star and became a RSG, and did not undergo full dredge up for some reason. One possible way would be if it were a low-metallicity star, but this is unlikely given the oxygen content of the host galaxy.
    \item It seems more likely that the progenitor was a mass gainer or merger in a binary system in such a way that it did not undergo substantial rotational or convective mixing.  We identify four merger models from \citet{glebbeek2013} that have surface abundances at the onset of He burning that could be consistent with our determinations for \object's progenitor. For that to hold, low rotational and convective mixing would also be required as to not substantially modify the model surface abundances until core collapse. These models have masses initial primary and secondary masses of $10+1$, $10+7$ and $20+2~\msun$. 
\end{enumerate}

Future observations of the environment of SN 2016bkv, once the SN is dim enough, may also contribute to determine the progenitor mass, in similar fashion as has been done for other core-collapse SNe \citep{murphy18}. In addition to further constraints on the age, color, and luminosity of the progenitor from population studies, a merger scenario would also predict that no companion should be detected. 

\section*{Acknowledgements}
We warmly thank the anonymous reviewer for a careful reading and constructive report, which greatly helped us to fix mistakes in the original submission and improve the quality of this paper. We appreciate that referee handled our manuscript in a courteous and professional manner.

\section*{Data availability}

The derived data generated in this research will be shared on reasonable request to the corresponding author.

\label{lastpage}
\end{document}